\documentclass{llncs}

\usepackage[T1]{fontenc}
\usepackage[latin1]{inputenc}
\usepackage{epsfig}
\usepackage{pst-all}
\usepackage{amssymb}

\newcommand{\coment}[1]{}

\newcommand{\siCond}{\mathtt{if\  }}

\newcommand{\chaineds}{\texttt{chained}}
\newcommand{\chainedsp}[1]{\texttt{chained}_{#1}}
\newcommand{\chained}[1]{\texttt{chained}(#1)}
\newcommand{\chainedp}[2]{\texttt{chained}_{#1}(#2)}

\newcommand{\letbar}[1]{\mbox{I\kern-0.23em#1}}
\newcommand{\nat}{\letbar N}

\newcommand{\letbarp}[1]{\mbox{I\kern-0.5em#1}}

\def\y{\;\wedge\;}
\def\o{\;\vee\;}
\newcommand{\eoc}{{\mathrm{otherwise\ }}}

\newcommand{\calO}{{\cal O}}

\newcommand{\bdfn}{\begin{definition} \begin{rm}}
\newcommand{\edfn}{\end{rm}$ $\qed \end{definition}}
\newcommand{\bthm}{\begin{theorem} \begin{rm}}
\newcommand{\ethm}{\end{rm}$ $\qed \end{theorem}}
\newcommand{\bprop}{\begin{proposition} \begin{rm}}
\newcommand{\eprop}{\end{rm}$ $\qed  \end{proposition}}
\newcommand{\bcor}{\begin{corollary}\begin{rm}}
\newcommand{\ecor}{\end{rm} \end{corollary}}
\newcommand{\blem}{\begin{lemma} \begin{rm}}
\newcommand{\elem}{\end{rm}$ $\qed  \end{lemma}}
\newcommand{\bfact}{\begin{fact} \begin{rm}}
\newcommand{\efact}{\end{rm} \end{fact}}
\newcommand{\bex}{\begin{example} \begin{rm}}
\newcommand{\eex}{\end{rm}$ $\qed  \end{example}}
\newcommand{\bprf}{\begin{proof}}
\newcommand{\eprf}{\end{proof}}

\begin{document}

\title{Some conditions implying {\it if P=NP then P=PSPACE\\
\footnotesize{(Work-in-progress draft)}}
}

\author{Ismael Rodríguez}
\institute{%
Dept. Sistemas Informáticos y Computación\\
Facultad de Inform\'atica\\
Universidad Complutense de Madrid, 28040 Madrid, Spain\\
e-mail: {\tt isrodrig@sip.ucm.es}\\[1.5em]
{\bf Last update: July 19, 2013} }

\maketitle
\begin{abstract} We identify a few conditions $X$ such that $$(P=NP \y X)
\;\Rightarrow\; P=PSPACE$$
\end{abstract}

{\bf Keywords:} Complexity, $P$, $NP$, $PSPACE$.

\section{Introduction}

It is known that $P \subseteq NP \subseteq PSPACE$, though neither
inclusion is known to be strict. In this draft we investigate whether
the condition

$$P=NP \Rightarrow P=PSPACE$$

\noindent holds. Proving this property would render $P=NP$ less
likely, as $P=PSPACE$ looks quite a strong assumption to most
researchers.

We identify a few sufficient conditions $X$ such that, if $X$ holds,
then we have $P=NP \;\Rightarrow\; P=PSPACE$. We prove that each of
them implies $P=NP \;\Rightarrow\; P=PSPACE$. We also discuss the
feasibility of each $X$, and we propose several ways to try to prove
each. Unfortunately, no proof of them is given in this draft.

This document is structured as follows. Our basic statement is
presented in the next section, and the feasibility of the first
proposed condition (in fact, a weaker version) is discussed in
Section~\ref{sec:feasibility}. Three methods to try to prove it are
discussed in Section~\ref{sec:ideas}. A different condition is
proposed in Section~\ref{sec:tqbf}, and a curious consequence of the
first proposed condition is introduced in Section~\ref{sec:curious}.
We present our conclusions in Section~\ref{sec:conclusions}.

\section{Basic statement}\label{sec:basic}

A Turing machine is defined by a tuple
$M=(S,\Sigma,\Gamma,s_1,s_A,\Delta)$ where $S$ is the set of states,
$\Sigma$ is the set of input symbols, $\Gamma$ is the set of tape
symbols ($\Sigma\subseteq \Gamma$), $s_1$ is the initial state, $s_A$
is the accepting state, and $\Delta$ is the transition set (only
deterministic Turing machines will be considered in this document).
We assume that $\sigma: S \cup \Gamma \rightarrow \{0,1\}^*$ returns
a codification of each symbol in $S$ and $\Gamma$ into
$\alpha=\ulcorner log_2(|S\cup \Gamma|)\urcorner$ bits. A
configuration of $M$ where $M$ is at state $s$, the (non-blank part
of) the tape is $c_1 \ldots c_v$, and the cursor is at the i-th
position, is denoted by the sequence of bits $\sigma(c_1) \ldots
\sigma(c_{i-1}) \sigma(s) \sigma(c_i) \ldots \sigma(c_v)$. Thus, the
initial configuration of $M$ for input $x=x_1 \ldots x_n$ is
$\sigma(s_1) \sigma(x_1) \ldots \sigma(x_n)$. We write $x
\Rightarrow_M y$ if $M$ can reach a configuration $y$ after $0$ or
more steps from its initial configuration $\sigma(s_1) \sigma(x_1)
\ldots \sigma(x_n)$ ($M$ does not necessarily terminate at $y$). This
initial configuration will be denoted simply by $\sigma(x)$.

Let $Q$ be a decision problem in $PSPACE$. There exists a
deterministic Turing Machine $M'=(S',\Sigma,\Gamma,s_1,s'_A,\Delta')$
such that $M'$ solves $Q$ in space $T(n)$ for some polynomial $T$. It
is trivial to construct a new Turing machine
$M=(S,\Sigma,\Gamma,s_1,s_A,\Delta)$ from $M'$ such that $M$ behaves
like $M'$, though it erases the tape after it reaches $s'_A$ and next
stops at the new accepting state $s_A$. In this way, we guarantee
that $M$ will have a single accepting configuration $\sigma(s_A)$.
Let us denote this accepting configuration by $c_A$.

Let us compute the $PSPACE$ problem $Q$ in an alternative way. Given
an input $x$, the question $x\in Q$? can be decided in the following
way:

\begin{itemize}
\item[1.] Construct from $M$ and $x$ a computation device capable
    of computing a finite boolean function (e.g. a propositional
    logic formula, a logic circuit, a finite automaton, a Turing
    machine, etc) behaving as the next boolean function
    $f_{M,x}$:

$$f_{M,x}(y) = \left\{\begin{array}{ll} 1 \;\;&\siCond x \Rightarrow_M y\\
    0& \eoc\end{array}\right.$$

\noindent Since $M$ operates in polynomial space $T(n)$, we may
define $f_{M,x}$ only for those configurations fitting into that
size. Thus the size of the finite domain of $f_{M,x}$ is
$2^{T(n)} \in \calO(2^{n^k})$ for some $k\in\nat$.

\item[2.] If $f_{M,x}(c_A)=1$ then answer {\it yes} else answer
    {\it no}.
\end{itemize}

Recall that $n$ is the size of $x=x_1\ldots x_n$. Let us introduce
two conditions (a) and (b) that will be used later to construct some
of the sufficient conditions $X$ mentioned in the introduction. It is
trivial to see that, if

\begin{itemize}
\item[(a)] there exists a polynomial $R$ such that step (1) can
    be performed in time $R(n)$, and
\item[(b)] there exists a polynomial $P$ such that the
    computation device constructed in~(1) runs in time $P(n)$ for
    all configurations whose size is bounded by $T(n)$,
\end{itemize}

\noindent then we can solve $Q$ for all $x$ in polynomial time by
performing (1) and next (2).

Note that the execution time of the constructed computation device
imposed by condition~(b) (i.e. $P(n)$) is defined in terms of $n$
(the size of $x$) rather than in terms of the size of the input of
that computation device (a configuration). However, the
configurations we are interested in have a size bounded by $T(n)$.
Hence, the execution time of the computation device is also
(indirectly) bounded by some polynomial of the size of its own
inputs.

It is also worth pointing out that $M$ is fix for $Q$, so the
complexity of this procedure to solve $Q$ does not depend on the size
of $M$ (thus, neither it depends on the size of $S$, $\Sigma$,
$\Gamma$, or $\Delta$).

Let us consider the following alternative condition:

\begin{itemize}
\item[(c)] there exist polynomials $D$ and $P$ such that, for all
    $x$, there {\it exists} a computation device computing
    $f_{M,x}$ whose representation size (e.g. the length of the
    propositional formula, the size of the logic circuit, the
    number of states of the AF or the TM, etc) is $D(n)$, {\it
    and} this computation device runs in time $P(n)$ for all
    configurations whose size is bounded by $T(n)$.
\end{itemize}

Let us note that the first requirement inside (c) (i.e. before {\it
and}) implies the second one for some kinds of computation devices
which always run in polynomial time with respect to the size of their
input (e.g. propositional formulas, finite automata). Indeed, for
computation devices running always in polynomial time with respect to
the size of their input, there must exist a polynomial $P$ such that
$P(n)$ steps are enough to process any input whose size is bounded
by~$T(n)$.

The important difference between considering conditions (a) and (b),
or considering the previous condition (c), is that, in the former
case, the computation device is required to be {\it computed} in
polynomial time, whereas in the latter case we just require that a
polynomial-size computation device {\it exists}.

In Section~\ref{sec:c} we will show that, if (c) holds and $P=NP$,
then (a) and (b) hold too. Since (a) and (b) allow us solving $Q$ by
performing steps (1) and (2) respectively, we infer that, if (c)
holds, then $Q$ can be solved in polynomial time provided that
$P=NP$. Let $A$ be the following condition:

$$A \;\equiv\; \mathrm{for\;all\;} M
\mathrm{\;running\;in\;some\;polynomial\;space\;} T(n), \mathrm{\;(c)\;holds}$$

If $A$ holds, then all problems in $PSPACE$ can be solved in
polynomial time provided that $P=NP$. That is,

$$(P=NP \y A) \Rightarrow P=PSPACE$$

Hence $A$ is one of the sufficient conditions $X$ mentioned in the
introduction. Sadly, no proof of condition $A$ is given in this
draft. The feasibility of condition $A$ will be discussed later, in
Section~\ref{sec:feasibility}. Possible approaches to try to prove (a
weaker version of) it will be presented in Section~\ref{sec:ideas}.

Next we prove that (c) and $P=NP$ imply (a) and (b).

\subsection{Proving that if (c) and $P=NP$ then (a) and
(b)}\label{sec:c}

Since (c) includes (b), (c) trivially implies (b). Let us show that
if $P=NP$ and (c) then (a). The following construction will, by the
way, let us discover a new sufficient condition $A'$ weaker than $A$.

How can we {\it construct}, in polynomial time, a model of a
(potentially) exponentially long execution of a Turing machine,
provided that some polynomial-size {\it representation} of that
(potentially) exponential execution {\it exists}?

Let us construct a function $\chaineds$ that, given a Turing machine
$M$, an input $x$ for that machine, a configuration $c$, and a
boolean function $f$ (more precisely, a computation device computing
$f$; in a notation abuse, both will be called $f$),
$\chained{M,x,f,c}$ returns $1$ if function $f$ fulfills, at
configuration $c$, some specific necessary condition to be a suitable
candidate to compute $f_{M,x}$. In particular, $\chained{M,x,f,c}$
will check whether $f$ is well ``chained'' to previous/next
configurations at configuration $c$ according to $M$, that is,
whether the answer of $f$ to $c$ is consistent with its answer to all
possible configurations that could happen immediately before $c$ in
$M$, as well as with the answer of $f$ to the configuration
immediately after $c$ according to $M$. If $f$ says that, when the
initial configuration is $x$, $M$ eventually traverses $c$
($f(c)=1$), then either $c=\sigma(x)$ (it is the initial
configuration) or there exists another configuration $c'$ such that
$f(c')=1$ and $c$ is reached from $c'$ in a single step of $M$. We
consider that $previous_M(c)$ denotes the set of configurations $c'$
such that, according to the transitions of $M$, the next
configuration after $c'$ would be $c$. Similarly, if $f$ says that
$c$ is reached, then either $c=c_A$ (the accepting configuration) or
$f$ must say that the next configuration of $M$ after $c$ is also
reached. We consider that $next_M(c)$ denotes the (single) next
configuration after $c$ in $M$.

Since by condition (c) we will assume that there exists a computation
device that computes $f_{M,x}$ in polynomial time with respect to $n$
(the size of $x$), any function $f$ spending a number of steps higher
than $P(n)$ for some polynomial $P$ can be considered irrelevant for
our purposes, so only computations of $f$ for up to $P(n)$ steps will
be checked in the execution of $\chaineds$. Actually, hereafter we
will speak about a $\chainedsp{P}$ function for each polynomial $P$,
rather than about a single $\chaineds$ function. In the following
definition of $\chainedsp{P}$, we will denote by $f(a)^{\leq
G(b)}\!\downarrow w$ that, in $G(b)$ steps or less, $f$ finishes its
computation for input $a$ and returns $w$.

According to these considerations, function $\chainedsp{P}$ is
defined as follows. Let us recall that $\sigma(x)$ is the initial
configuration of $M$.

{\footnotesize
$$\chainedp{P}{M,x,f,c}=\left\{\begin{array}{ll}1 & \siCond \begin{array}[t]{l} 
\left(\begin{array}{l}f(c)^{\leq P(n)}\!\downarrow 0 \y 
c \neq \sigma(x)
\end{array}\right)\\\o\\
\left(\begin{array}{l}f(c)^{\leq P(n)}\!\downarrow 1 \y\\
(c\neq \sigma(x) \rightarrow \exists\;c'\in previous_M(c): f(c')^{\leq P(n)}\!\downarrow 1) \y\\
(c\neq c_A \rightarrow f(next_M(c))^{\leq P(n)}\!\downarrow 1)\end{array}\right)
\end{array}\\0 & \eoc\end{array}\right.$$}

We can see that $\chainedsp{P}$ operates in polynomial time with
respect to the size of its input (a tuple $(M,x,f,c)$).
All calls to function $f$ are executed for a number of times ($n$)
which is the size of a part of the input tuple ($x$). Besides, let us
note that computing $previous_M(c)$ just consists in traversing
backwards all transitions that could reach configuration $c$
according to $M$, and the amount of these transitions is constant
with respect to $n$ because the size $M$ (and the size of its
transition set $\Delta$) is constant with respect to $n$. Also,
$next_M(c)$ returns a single configuration and is trivial to compute.

We will use function $\chainedsp{P}$ to find a sufficient (but not
necessary) condition to detect that $f\neq f_{M,x}$: if
$\chainedp{P}{M,x,f,c}=0$ for some $c$ whose size does not exceed
$T(n)$ then clearly $f$ does not compute $f_{M,x}$, because $f$ does
not {\it chain} correctly its behavior for some previous/next
configuration.
Let us note that, even if $\chainedp{P}{M,x,f,c}=1$ for all of those
$c$, we could have $f\neq f_{M,x}$. In particular, there could exist
some configurations $c_1,\ldots,c_z$ such that
$f(c_1)=1,\ldots,f(c_z)=1$ but $M$ does {\it not} eventually traverse
any of these configurations when it receives input $x$. Let
$c^*\in\{c_1,\ldots,c_z\}$, so $c^*$ is not eventually traversed when
$M$ receives $x$ but we have $f(c^*)=1$. Note that, if $f$ returns
$1$ for some configuration, it is also required to return $1$ for
some immediately previous configuration as well as for some
immediately subsequent configuration. By going backwards (according
to $M$ transitions) from $c^*$ through some configurations $c'$ with
$f(c')=1$, we will never reach a configuration $c''$ that is
eventually traversed by $M$ from configuration $\sigma(x)$: If
$f(c'')=1$ then $f(next(c''))=1$ and so on, so all configurations
included in our previous backwards traversal would indeed be
traversed by $M$ (included $c^*$, which is contradictory). Thus,
$f(c^*)=1$ is possible only if $c^*$ is part of a {\it cycle} of
consecutive configurations such that {\it none} of them is actually
reached by $M$ from $\sigma(x)$ (but $f$ returns $1$ for all of
them).
Let us note that $M$ {\it stops} at $c_A$ (i.e. the single acceptance
configuration), so a cycle cannot include $c_A$, and $M$ cannot
escape from a cycle because it is deterministic. That is, $f(c_A)=1$
iff $M$ reaches $c_A$ from $x$ indeed.

This means that any computation device $f$ such that
$\chainedp{P}{M,x,f,c}=1$ for all configurations $c$ whose size is
under $T(n)$ (recall that $T(n)$ is an upper bound of the space used
by $M$ for input $x$) can be used in step (2) of the algorithm shown
in the previous section: Even if $f$ includes spurious cycles like
those commented before, $M$ accepts $x$ iff $f(c_A)=1$. Thus,
condition (c) given in the previous section can be relaxed. We do not
need that the function having the required size and computing in the
mentioned time is $f_{M,x}$ indeed. We can also use any function $f$
fulfilling $\chainedp{P}{M,x,f,c}=1$ for all $c$ within the required
size, even in $f\neq f_{M,x}$. This enables the following alternative
definition of the condition (c) mentioned in the previous section:

\begin{itemize}
\item[(c')] there exists polynomials $D$ and $P$ such that, for
    all $x$, there {\it exists} a computation device $f$ such
    that its size is equal to or lower than $D(n)$; we have
    $\chainedp{P}{M,x,f,c}=1$ for all $c$ whose size is at most
    $T(n)$; and $f$ runs in time $P(n)$ for all $c$ within that
    size.
\end{itemize}

This alternative requirement (c') enables the following alternative
condition~$A'$:

$$A' \;\equiv\; \mathrm{for\;all\;} M
\mathrm{\;running\;in\;some\;polynomial\;space\;} T(n), \mathrm{\;(c')\;holds}$$

In the rest of the section we will show that

$$(P=NP \y A') \Rightarrow P=PSPACE$$

It is easy to see that $A$ implies $A'$ (as (c) implies (c'), because
we can take $f=f_{M,x}$), so proving the previous statement will also
prove the original statement given in the previous section (the one
concerning $A$ instead of $A'$).

Before going on with the proof, we briefly introduce two alternative
constructions which could also be used:

\begin{itemize}
\item[(I)] It is easy to modify our construction so that only
    functions {\it without} spurious cycles are allowed by
    $\chainedsp{P}$. We just have to consider that configurations
    also include an additional numeric parameter denoting the
    current {\it number of step} in the execution of $M$. Note
    that execution step numbers can be represented in polynomial
    size: since $M$ runs in polynomial space $T(n)$, its number
    of execution steps for any input is at most exponential with
    $T(n)$ provided that $M$ does not repeat any configuration
    and loops forever (the number of possible configurations is
    exponential with $T(n)$), so just a polynomial number of bits
    is required to denote any execution step number. Given a
    configuration attached to number $r$, the previous
    configuration is necessarily attached to number $r-1$, and
    the next one to $r+1$. This way, cyclic behaviors would not
    be allowed by $\chainedsp{P}$.
\item[(II)] Let us consider again that spurious cycles are
    allowed. Without loss of generality, we could modify $M$ so
    that, when it reaches $c_A$, it does not stop but it writes
    the initial configuration on the tape and goes back to state
    $s_1$. That is, it would go back from $c_A$ to $\sigma(x)$,
    so a path from the initial configuration towards acceptation
    would {\it also} be within a cycle. In this case, only
    functions $f$ that represent a set of execution cycles (and
    nothing else) could fulfill $\chainedp{P}{M,x,f,c}=1$ for all
    $c$. Still, any function $f$ fulfilling that condition would
    also be valid for step~(2) of our algorithm: since $c_A$
    leads to $\sigma(x)$, $c_A$ can belong to a cycle only if
    $\sigma(x)$ {\it also} leads to $c_A$. Thus we would also
    have $f(c_A)=1$ iff $M$ accepts $x$.
\end{itemize}

Next we resume the proof. Let us consider the following set:

$$G_{P,T}=\left\{(M,x,f)\left| \begin{array}{l}
\exists\;c: (\mathrm{the\;representation\;size\;of\;} c \mathrm{\;is} \leq T(n) \y\\
\;\;\;\;\;\;\;\chainedp{P}{M,x,f,c}=0) \end{array}\right.\right\}$$

$G_{P,T}$ consists of all triples $(M,x,f)$ such that $f$ ``breaks''
the execution chain at some point (i.e. we have
$\chainedp{P}{M,x,f,c}=0$ for some $c$ whose size is under the
required threshold $T(n)$).

Let us show that $G_{P,T}\in NP$. Given a triple $(M,x,f)$, a
non-deterministic Turing machine (NDTM) can determine in polynomial
time whether $(M,x,f)\in G_{P,T}$ holds as follows. First, it
non-deterministically generates any $c$ whose representation size is
under $T(n)$. Next, it deterministically checks
whether we have $\chainedp{P}{M,x,f,c}=0$, which requires polynomial
time. Thus, this non-deterministic algorithm determines in polynomial
time whether $(M,x,f)\in G_{P,T}$ holds.

Since we are assuming $P=NP$, we have $NP=co\!-\!NP$, so the set

$$\overline{G_{P,T}}=\left\{(M,x,f)\left| \begin{array}{l}
\forall\;c: (\mathrm{the\;representation\;size\;of\;} c \mathrm{\;is} \leq T(n) \rightarrow\\
\;\;\;\;\;\;\;\chainedp{P}{M,x,f,c}=1) \end{array}\right.\right\}$$

\noindent also belongs to $NP$. Since $P=NP$, we deduce
$\overline{G_{P,T}}\in P$.

Let us consider the following set:

$$H_{P,T}=\left\{(M,x,f,s)\left| \begin{array}{l}
(M,x,f) \in \overline{G_{P,T}} \y\\
\mathrm{the\;bit\;sequence\;representing\;} f \mathrm{\;is\;}\\
\;\;\;\;\mathrm{lexicographically\;lower\;than\;the\;bit\;sequence\;} s
\end{array}\right.\right\}$$

We have $H_{P,T}\in P$ because $\overline{G_{P,T}}\in P$ and the
second condition can also be checked in polynomial time.

Thus, the set

$$W_{P,D,T}=\left\{(M,x,s)\left| \begin{array}{l}
\begin{array}{ll}\exists\;f: & (\mathrm{the\;representation\;size\;of\;} f \mathrm{\;is} \leq D(n) \y \\
& (M,x,f,s) \in H_{P,T})
\end{array}
\end{array}\right.\right\}$$

\noindent belongs to $NP$: A NDTM just has to non-deterministically
generate any $f$ whose representation size is under $D(n)$ and next
deterministically check whether $(M,x,f,s) \in H_{P,T}$ holds, which
can be done in polynomial time.

Since $P=NP$, we also have $W_{P,D,T}\in P$. Thus, the problem of
checking whether there exists $f$ such that its size is at most
$D(n)$, $f$ is lexicographically lower than a given bit string $s$,
and $f$ is valid for our purposes (i.e. $(M,x,f)\in
\overline{G_{P,T}}$) can be determined in polynomial time.

According to that property of $W_{P,D,T}$, we conclude that the step
(1) of the algorithm proposed in the previous section can be solved
in polynomial time by operating as follows. Given $M$ and $x$, we
perform a binary search to find some $f$ fulfilling $(M,x,f)\in
\overline{G_{P,T}}$, and this is done by performing several calls to
the procedure solving $(M,x,s)\in W_{P,D,T}$ with different values of
$s$. We start by setting $s$ to the midpoint of the set of possible
values of $f$, and next we iteratively call $(M,x,s)\in W_{P,D,T}$
and set $s$ to the midpoint of one half of the range or another,
depending on whether $(M,x,s)\in W_{P,D,T}$ holds or not. The number
of calls to the procedure checking $(M,x,s)\in W_{P,D,T}$ is
polynomial and each call requires polynomial time, so function $f$ is
found in polynomial time --provided that it exists.

By assumption (c'), $f$ actually exists, so step (1) succeeds.

After we get $f$ in step (1), we check whether $f(c_A)=1$ holds,
which tells us if $x\in Q$ in polynomial time.

Thus, the previous procedure allows us to decide, in polynomial time,
any problem in $PSPACE$ provided that the required conditions hold.

\section{Discussion on the feasibility of condition $A'$}\label{sec:feasibility}

In this section we discuss the feasibility of conditions $A$ and
$A'$. Since $A'$ is weaker than $A$, only $A'$ will be mentioned
hereafter. Arguments given in this section in favor or against $A'$
will be arguable, sometimes vague, and generally weak.

First, let us note that $A'$ would be trivially met if {\it only one}
of the two conditions imposed in $(c')$ were required. On the one
hand, we can construct a {\it constant-size} computation device $f$
that tells us whether $M$ traverses any configuration $c$ when it
receives $x$, though it can take exponential time to decide: We just
have to simulate $M$ and check whether $c$ is traversed. On the other
hand, we can construct an exponential-size computation device $f$
that tells us, in {\it polynomial time}, whether $M$ traverses $c$:
$f$ is constructed as a binary decision tree where each each path
from the root to a leaf denotes each possible configuration of size
$T(n)$, and leaves say yes or not depending on whether that
configuration is traversed by $M$ from $x$. An execution of $f$
consists in following the path representing the selected
configuration from the root to a leaf ($T(n)$ steps) and providing
the answer in the leaf. Thus, we can trivially fulfill either one
polynomial boundary imposed by $(c')$ alone, or the other one alone
(moreover, the first one can be made {\it constant}, far under the
polynomial requirement). The following question arises: Can both
polynomial limits be met {\it together}?

Let us note that most of boolean functions cannot be represented by a
sequence of bits of polynomial size with respect to the size of the
input~\cite{sha49}, no matter how we codify these sequences of bits
(or whether they represent Turing Machines, Finite Automata, formulas
of propositional logic, logic circuits, etc). If we consider boolean
functions that receive $n$ bits and return $1$ bit, then there exist
$2^{2^n}$ of them, and uniquely identifying them with a binary code
requires $log_2(2^{2^n})=2^n$ bits if all of them are identified with
the same number of bits. Alternatively, we could assign shorter codes
to some of them, but just a {\it few}. For instance, we can codify
$2^{n^k}$ functions with only $log_2(2^{n^k})=n^k$ bits, but then the
ratio of boolean functions we can represent with such polynomial size
would be only $\frac{2^{n^k}}{2^{2^n}}$. That is, only a {\it few}
functions can be assigned polynomial-size codes.

This looks to be bad news for property $A'$, which requires
representing, with a {\it polynomial-size} code, the boolean function
identifying all configurations traversed by a Turing machine for a
given input. However, let us note that the set of traversed
configurations is not {\it any} kind of set, but it is constrained by
the Turing machine under consideration. In particular, traversed
configurations are necessarily linked with each other as the machine
transitions define: if a configuration is reached, then some previous
one, as well as the next one, are reached. This requirement strongly
constrains the form of boolean functions under consideration.

However, the most obvious constraint imposed by the Turing machine is
the fact that the size of the Turing machine is {\it constant} with
respect to the size of the input $x$.
Let $p,n\in \nat$. Let $J$ be the set of all pairs $(M,x)$ where $M$
is a Turing machine with $p$ states and $x$ is an input with $n$
bits. Can the set of boolean functions identifying all configurations
traversed by all pairs $(M,x)\in J$ contain {\it all} boolean
functions from $n$ bits into $1$ bit (i.e. $2^{2^n}$ functions)?
Though it might be possible if $p$ is much higher than $n$, we just
have to consider a higher $n$ to make it impossible. Recall that
property $A'$ is defined in asymptotic terms with respect to $n$, so
let us consider it in that way. The size of $M$ is constant with $n$
so, in asymptotic terms, a pair $(M,x)\in J$ can represent only up to
$\calO(2^n)$ different boolean functions (there are $2^n$ possible
inputs $x$ with size $n$, and at most we have a different function
for each of them, so up to $2^n$ different functions can be denoted).
That is, for inputs of size $n$, there are $2^n$ possible boolean
functions to consider in asymptotic terms, rather than $2^{2^n}$. If
there are {\it only} $2^n$ functions to be represented, then all of
them could be identified with some polynomial-size code according to
some codification.
%
%
The intuitive idea is that the ``simplicity'' of the set of
configurations traversed by a Turing machine arises asymptotically,
when we consider big inputs.
%

Unfortunately, there are several incompatible notions of simplicity:
something that is simple for a given representation system might be
complex for another. For instance, it is known that the parity
function cannot be computed with a logic circuit with $\calO(1)$
depth and polynomial size~\cite{has87}. This means, in particular,
that the parity cannot be computed by a polynomial-size DNF or CNF,
which could be surprising at a first glance due to the simplicity of
such function. However, a trivial Finite Automaton with just two
states computes the parity. Thus, several notions of simplicity could
be mutually incompatible, and the kind of simplicity required by
property $A'$ could not be met, even asymptotically.

Another possible argument in this discussion is that many classic
Computability and Complexity results directly or indirectly show us
that, if you want to know what a Turing machine will do in the next
$n$ steps, then in general you have to execute/simulate these $n$
steps, there are no shortcuts. For instance, discovering that a
program will {\it not} halt essentially requires simulating it
forever. However, the algorithm (1)-(2) given in
Section~\ref{sec:basic} would allow us to know, in polynomial time,
what a Turing machine will do in (potentially) exponential time.
So, if $A'$ were true, then we would have a way to ``cheat'' and
infer, in $k$ steps, what will happen in $n>k$ steps.

Let us note that our construction does {\it not} prove that we can
cheat like this provided that $A'$ holds. It just proves that we
could cheat like this provided that we have $A'$ {\it and} $P=NP$. If
$P\neq NP$ were true (as most researchers think) then such a way of
cheating would be impossible, which matches our natural intuition
indeed.

Let us compare property $A'$ with some requirements related to
Circuit Complexity classes. Typically, Circuit Complexity classes
require that there exists a Turing machine that, given the size of
the input to be solved, constructs the finite circuit solving the
problem for all inputs of each size. Thus, if we want to solve a
problem for an input of size $n$, first our Turing machine produces a
circuit that is suitable for inputs of that size, and next the
circuit is used to solve the problem. If polynomial-time or logspace
uniformity is required, then the Turing machine constructing these
circuits is required to construct that circuit in {\it polynomial}
time or {\it logarithmic} space, respectively. However, our property
$A'$ does not require that a computation device $f$ fulfilling
$\chainedp{P}{M,x,f,c}$ for all $c$ of some size can be {\it
computed} in some required time or space. It just requires that the
computation device (Turing machine, propositional formula, etc) {\it
exists}, which is quite a weaker requirement.

\section{Discussion: ideas to try to prove $A'$}\label{sec:ideas}

In this section we discuss possible ways to prove $A'$. All of these
ideas are very speculative.

We consider the following possibilities:

\begin{itemize}
\item[(a)] The definition of $\chainedsp{P}$ shows the kind of
    requirement that has to be preserved by $f$ in order to be
    correct. For each Turing machine $M$, we can define this
    requirement as a {\it functional equation} where the
    definition of function $f$ is given as a propositional logic
    expression depending on the value of {\it itself} for other
    parameters (those denoting {\it previous} and {\it next}
    configurations according to $M$). Could fix point evaluation
    techniques be used to find the particular form of any
    function $f$ fulfilling this kind of functional equations,
    and show that this form will always be able to be simplified
    up to a polynomial size? If so, property $A'$ would hold, as
    propositional logic formulas can always be evaluated into
    true or false in polynomial time with respect to their size.
\item[(b)] Turing machines can be easily simulated by
    one-dimension cellular automata (CA). The state of each cell
    in a CA at time $t+1$ is defined by a boolean function
    depending on a constant number of neighbor cells at time $t$,
    and all cells have the same boolean function defined in terms
    of its local neighbors. If we want to simulate a TM with a
    CA, we may construct a specific CA for the TM, or we may use
    a universal CA to simulate any TM (e.g. the universal
    automaton $110$~\cite{coo04,nw06}, where the boolean function
    of all cells depends only on the current value of the cell
    and its two adjacent neighbors). Let us consider a CA that
    simulates a TM working in $n^k$ space. In each transition of
    the CA from time $t$ to time $t+1$, some function $f$, from
    $n^k$ bits into $n^k$ bits, is applied to all cells. Thus, if
    $x$ is the initial state of all cells, then the state of the
    CA after simulating $z$ TM steps is $f \circ f \circ \ldots
    \circ f (x) = f^z(x)$. Moreover, the function we apply at
    each step, $f$, is the result of applying the same
    $\calO(1)$-size local boolean function to all cells in the
    tape, so the computation of a CA is the result of applying
    some simple $\calO(1)$-size logic circuit symmetrically both
    in space and time. Could this peculiarity imply that the set
    consisting of $x$, $f(x)$, $f^2(x)$, \ldots, $f^{2^{n^k}}(x)$
    (i.e. the set of all configurations traversed by the original
    TM for input $x$) can always be simplified into a
    polynomial-size representation? If so, $A'$ would hold.
\item[(c)] If we view function $f$ as an $n^k$-cube where
    vertexes denote configurations, then transitions from a
    configuration to another can be viewed as vectors in this
    $n^k$-cube space. All vertexes have a single outgoing vector
    (i.e. they lead $M$ to another single configuration), so we
    can view the whole picture as a vector field. Since there
    exist $\calO(1)$ transitions in the transition set $\Delta$
    of $M$, all vectors in the $n^k$-cube fit into one of
    $\calO(1)$ kinds. Let us suppose that a transition of $M$
    says how $M$ must change when $M$ is at state $s_4$ and the
    tape symbol pointed by the cursor is $1$. Let us suppose that
    configuration substring $1011$ means ``the state is $s_4$ and
    the cursor is here'' whereas $0100$ means ``the symbol here
    is $1$.'' According to the notation for configurations
    proposed at the beginning of Section~\ref{sec:basic}, the
    cursor is located at the symbol to the right of the
    configuration substring denoting the state, so this
    transition will be triggered for any configuration containing
    the $10110100$ substring. In particular, any configuration
    where this substring is contained (at any location) will
    react in the {\it same} way.

    Hence we have this pattern of ``simplicity'' or
    ``repetitiveness'' in the vector field. Could it make the
    $n^k$-cube have a {\it polynomial} number of $n^{k'}$-cubes,
    with $k'<k$, such that all vertexes denoting traversed
    configurations (or configurations within spurious cycles, as
    explained in Section~\ref{sec:c}) are included in one of
    them? Let us note that each $n^{k'}$-cube within the
    $n^k$-cube can be represented by a conjunctive term. For
    instance, the cube where $x_1$ is true, $x_4$ is false, and
    $x_7$ is true is denoted by the term $x_1 \y ¬x_4 \y x_7$.
    This term denotes all vertexes fulfilling that condition
    regardless of their values for $x_2$, $x_3$, $x_5$, etc. A
    propositional formula returning $1$ exactly for all
    configurations belonging to some of these $n^{k'}$-cubes
    could be constructed just by forming the disjunction of all
    of those terms. That is, the resulting DNF would compute $f$.
    If the number of conjunctive terms in $f$ (i.e. the number of
    $n^{k'}$-cubes) is polynomial, then this DNF has polynomial
    size (and, trivially, $f$ could be computed in polynomial
    time).

    The fact that we can choose whether each spurious cycle is
    included in $f$ or not (they are unnecessary but harmless)
    adds some flexibility to pick these $n^{k'}$-cubes. Moreover,
    some spurious {\it transitions} (i.e. transitions that do not
    change the behavior of $M$ for $x$) could be artificially
    added to add even more flexibility.
\end{itemize}

So far we have not considered any approach to prove $A'$ where $f$ is
represented by a Turing machine running in polynomial time (rather
than by logic circuits or propositional formulas, as in ideas (a),
(b), and (c)). Reasoning about what can be done (or not) with Turing
machines whose size (e.g. number of states) is under some given
threshold is typically hard. Maybe some diagonalization argument
could work here?

The {\it Kolmogorov complexity} studies problems in terms of the size
of the programs that solve them, and some {\it time-bounded}
Kolmogorov complexity notions have been proposed, including
polynomial-time ones (see e.g.~\cite{for04}). Unfortunately, to the
best of our knowledge there is no result concerning, specifically,
the Kolmogorov complexity of polynomial-time Turing machines that
identify the finite set of configurations traversed by some
polynomial-space Turing machine for some input. Anyway, we hope a
deeper study of the literature on this field could provide some hints
for facing our problem.

\section{A different approach: dealing with a $PSPACE$-complete
problem}\label{sec:tqbf}

In this section we consider a very different approach to find a
sufficient condition for $P=NP \Rightarrow P=PSPACE$. Rather than
generically considering any PSPACE problem, we will focus on a
specific $PSPACE$-complete problem. In particular, we will show a
property such that, if it holds, then some $PSPACE$-complete problem
could be solved in polynomial time provided that $P=NP$ holds. Since
all $PSPACE$ problems can be polynomially reduced into any
$PSPACE$-complete problem, this would prove $P=PSPACE$.

Let us consider the TQBF problem ({\it true quantified boolean
function} problem). This $PSPACE$-complete decision problem is
defined as follows. Let $\varphi$ be a propositional logic formula
over propositional symbols $x_1,\ldots,x_n$. Let us consider the
following expression:

$$Q_1 x_1 Q_2 x_2 Q_3 x_3 Q_4 x_4 \ldots Q_k x_k \;\varphi$$

\noindent where $Q_i\in \{\forall,\exists\}$ and quantifiers
$\forall$ and $\exists$ strictly alternate through the sequence
$Q_1,\ldots,Q_k$. If we have a fully quantified expression (i.e.
there are no free variables) then we will say that we have a QBF
formula.

TQBF is the set of QBF expressions that evaluate to $\top$. Thus,
solving TQBF consists in checking whether the given expression is
equivalent to $\top$ or $\bot$. Since all variables are quantified,
all expressions must be equivalent to either $\top$ or $\bot$.

A straightforward way to evaluate expressions like those consists in
iteratively getting rid of quantifiers, from the last one to the
first one, by performing the following replacements:

$$\begin{array}{c}\forall x\; \varphi \equiv \varphi[x/\top] \y
\varphi[x/\bot]\\[0.5em]
\exists x\; \varphi \equiv \varphi[x/\top] \o \varphi[x/\bot]\end{array}$$

This method doubles the size of the resulting formula at each step,
so the formula grows exponentially. However, when the last quantifier
is removed, the expression has no variables and it is equivalent to
either $\top$ or $\bot$.

Let $\gamma$ be a QBF expression and let $n$ be its size. Let
$q_\gamma < n$ denote the number of quantifiers in $\gamma$. Let us
suppose that we evaluate $\gamma$ as proposed before, though the
propositional part of the QBF resulting after each replacement is
simplified up to its minimal form before performing the next
replacement. We denote by $s_1^\gamma,\ldots,s_{q_\gamma}^\gamma$ the
sizes of the (simplified) propositional parts of the formula after
the first replacement, the second one, \ldots, and the $q_\gamma-$th
one, respectively.

We introduce the following condition:

$$B \equiv \begin{array}[t]{l}
\mathrm{there\;exists\;a\;polynomial\;} P \mathrm{\;such\;that,}\\ \mathrm{for\;all\;QBF\;} \gamma,\; s_i^\gamma \leq P(n)
\mathrm{\;for\;all\;} 1\leq i\leq q_\gamma \end{array}$$


Let us show that:

$$(P=NP \y B) \Rightarrow P=PSPACE$$

If property $B$ holds, then we can infer whether QBF is $\top$ or
$\bot$ in polynomial time provided that we can simplify propositional
formulas in polynomial time. By simplifying the QBF expression after
getting rid of each quantifier, the size of the resulting formula
always remains polynomial. This operation has to be repeated a linear
number of times, because there is a linear number of quantifiers in
the formula ($q_\gamma$ cannot be higher than $n$). Thus the whole
process takes polynomial time.

Let us suppose $P=NP$. Simplifying boolean functions is an
$\Sigma_2^p$ problem (in fact, it is $\Sigma_2^p$-complete under
Turing reductions~\cite{bu11}). Since the whole polynomial hierarchy
would collapse if $P=NP$, we know that $P=NP$ implies that
propositional formulas can be simplified in polynomial time.

We infer $(P=NP \y B) \Rightarrow P=PSPACE$.

Do we have any intuitive reason to suspect that $B$ holds? Rather
than addressing this issue from a general point of view, let us
consider a particular case of QBF.

Any QBF instance can be polynomially reduced into an equivalent QBF
where the propositional part ($\varphi$) is given, in particular, in
CNF~\cite{sip06}. Let us see what happens in the evaluation of a QBF
when $\varphi$ is given in CNF.

Given a set $S$ of clauses, we denote by $S_{x}$ the subset of
clauses of $S$ where the $x$ literal appears, by $S_{¬x}$ the subset
of clauses of $S$ where $¬x$ appears, and by $S_{[x]}$ the subset of
clauses of $S$ where neither $x$ nor $¬x$ appear. If $c$ is a clause,
we denote by $c_{[x]}$ the result of removing any appearance of $x$
at clause $c$. For instance, $(x_1 \o ¬x_2 \o x_3)_{[x_2]}=(x_1 \o
x_3)$.

Let $S$ be the set of clauses in $\varphi$. We have:

$$\begin{array}{ll}\exists\;x\; \varphi \equiv &
\varphi[x/\top] \o \varphi[x/\bot] \equiv \\
& (\bigwedge_{c\in S_{[x]}} c \y \bigwedge_{c\in S_{¬x}} c_{[x]}) \o
(\bigwedge_{c\in S_{[x]}} c \y \bigwedge_{c\in S_{x}} c_{[x]}) \equiv\\
& \bigwedge_{c\in S_{[x]}} c \y (\bigwedge_{c\in S_{¬x}} c_{[x]} \o \bigwedge_{c\in S_{x}} c_{[x]})
\end{array}$$

$$\begin{array}{ll}\forall\;x\; \varphi \equiv &
\varphi[x/\top] \y \varphi[x/\bot] \equiv \\
& (\bigwedge_{c\in S_{[x]}} c \y \bigwedge_{c\in S_{¬x}} c_{[x]}) \y
(\bigwedge_{c\in S_{[x]}} c \y \bigwedge_{c\in S_{x}} c_{[x]}) \equiv\\
& \bigwedge_{c\in S_{[x]}} c \y \bigwedge_{c\in S_{¬x}} c_{[x]} \y \bigwedge_{c\in S_{x}} c_{[x]}
\end{array}$$

Let us note that the size of $\forall\;x\; \varphi$ is not lower than
the size of the expression it turns into, that is $\bigwedge_{c\in
S_{[x]}} c \y \bigwedge_{c\in S_{x}} c_{[x]} \y \bigwedge_{c\in
S_{¬x}} c_{[x]}$. Even if the same number of clauses appear, some of
them could be shorter. Thus, the only risk to increase the formula
size is due to the replacement of $\exists\;x\; \varphi$ by
$\bigwedge_{c\in S_{[x]}} c \y (\bigwedge_{c\in S_{x}} c_{[x]} \o
\bigwedge_{c\in S_{¬x}} c_{[x]})$. If we apply the distributive law
to subformula $(\bigwedge_{c\in S_{x}} c_{[x]} \o \bigwedge_{c\in
S_{¬x}} c_{[x]})$ to convert it back into a CNF, in general the
number of clauses grows quadratically with respect to the size of
this subformula.

The new clauses created by the application of the distributive law to
that subformula (each clause has up to double size) cannot contain a
literal and its negated literal: in this case, the whole clause is
trivially true and can be eliminated from the conjunction of clauses.
Also, after all new clauses are deployed, we could match some pairs
of clauses following the form $(\varphi_1 \o x)$ and $(\varphi_1 \o
¬x)$. In this case, both of them could be replaced by a single clause
$\varphi_1$.

It is easy to create QBF expressions where, during the elimination of
the first quantifiers, the quadratic expansion due to the application
of the distributive law at $\exists$ eliminations clearly surpasses
the simplifications produced later in $\exists$ eliminations or in
all $\forall$ eliminations. If this tendency is kept during many of
these eliminations, then the size of the QBF will trivially grow
exponentially. However, in some examples developed by hand we
observed that the number of available simplifications also grows fast
during the elimination of the first quantifiers. Could it happen that
the ``reduction force'' begins to surpass the ``expansion force''
{\it before} the formula has reached an exponential growth? Let us
note that simplifications always win: eventually, the formula will be
either $\top$ or $\bot$. However, {\it when} do they begin to win? Do
they do so before the expression can reach an exponential size?

If $P=NP$ then propositional formulas can be simplified in polynomial
time. This also applies to restricted subsets of propositional
formulas, such as CNF and DNF. Thus, if the simplification force
surpasses the expansion force before the QBF grows exponentially
then, by iteratively getting rid of each quantifier and next
simplifying the resulting formula (in polynomial time due to $P=NP$),
we could solve the TQBF problem in polynomial time, so $P=PSPACE$.

\subsection{Possible methods to study property $B$}\label{sec:expB}

Performing computer experiments to empirically check property $A'$,
given in Section~\ref{sec:c}, looks like a quite difficult goal.
First, it requires picking Turing machines working in polynomial
space (and so perhaps in exponential time) and run them for some
inputs. Collecting all configurations traversed by one of these
Turing machines for some input requires collecting a potentially
exponential number of configurations. After these configurations are
collected, we have to check whether a polynomial-size computation
device representing them, and running in polynomial time, exists. For
instance, we can represent them by a propositional formula, which
obviously runs in polynomial time with respect to its size. In this
case, finding a polynomial-size representation consists in
simplifying a propositional formula that {\it extensionally} covers
all traversed configurations. Let us note that the latter can be
constructed just as a DNF containing a conjunctive term for each
configuration traversed by the Turing machine. Simplifying that
extensional expression consists in solving an exponential-time
problem (the best known simplification algorithms are exponential)
for a DNF of exponential size. So the whole method would be doubly
exponential! The difficulty is even higher if we also wish to check
if the simplification works better under the presence/absence of some
spurious cycles or some spurious harmless additional transitions.

On the contrary, performing similar experiments for empirically
checking property $B$ looks much easier. The method proposed in
Section~\ref{sec:tqbf} to solve TQBF consists in applying a new
simplification after we get rid of each quantifier. Since the number
of quantifiers is linear with the size of the QBF formula, an
exponential problem (simplification) has to be solved a {\it linear}
number of times. This makes this goal easier, so bigger instances
could be analyzed. Thus, our first step will be performing an
experiment to empirically check B for a large number of cases.

If experiments show that B holds for many cases, we would try to {\it
prove} it. 
A possible method to do so would be developing a formal model of the
expansion and contraction forces of the QBF during the process where
quantifiers are eliminated, and finding out a kind of invariant of
the size of the simplified QBF along time.


\section{A picturesque consequence of $A'$}\label{sec:curious}

Let us briefly present a curious consequence of property $A'$.

In Section~\ref{sec:c} we saw that if $A'$ holds then, for all TM $M$
running in polynomial space and all input $x$, there exists a
polynomial-size computation device that is {\it well chained} with
respect to $M$ and runs in polynomial time. That is, $A'$ enables the
existence of a {\it small} and {\it fast} well-chained computation
device for $M$ {\it and each specific input $x$}. Next we show that
there also exists a small and fast well chained computation device
for $M$ and {\it all inputs} of some size, that is, one
simultaneously dealing with all inputs of some size.

Let $M$ be a Turing machine operating in polynomial space. For the
sake of simplicity, let us assume that inputs of $M$ are binary
sequences, i.e. strings in $\{0,1\}^*$. Let $s_A$ be the (single)
accepting state of $M$. Without loss of generality, we also assume
that it also has a single rejecting state $s_R$ (if not, we can
modify $M$ so that all combinations of state and symbol tape where
$M$ would stop actually lead to $s_R$, where next it really stops).

Let us modify $M$ to make it process all inputs of some size in a
{\it single} execution. At the initial state of $M$, we introduce new
transitions to make it immediately copy the input somewhere else in
the tape (we assume that the original machine $M$ never reaches the
area where it is copied). Next, the machine goes back to the
beginning of the original input and reaches the original initial
state~$s_1$.

Besides, we also add other transitions so that, when $M$ reaches
$s_A$, it erases all the tape except the input copy mentioned in the
previous paragraph, goes to the left boundary of that copy, and
reaches a new state $s'_A$. Next, it adds $1$ to that sequence,
copies the resulting sequence into the area where $M$ originally
received its input, points to the first symbol of that new copy, and
reaches $s_1$.

We do similarly with the rejecting state $s_R$: we add new
transitions so that, when $M$ reaches $s_R$, it also erases all the
tape except the input copy, reaches the left boundary of the copy,
and then reaches a new state $s'_R$. Next it adds $1$ to the input
copy, copies it at the original working area of $M$, points to the
first symbol of that new copy, and reaches state $s_1$.

Let $m\in\nat$. We also modify $M$ so that, when it adds $1$ to the
input copy, it checks whether the copy is higher than $m$. If so
then, instead of copying it and going back to $s_1$ again, it just
stops.

Let $M'$ be the resulting new Turing machine after introducing all of
these changes. We can see that $M'$ iteratively simulates $M$ for all
inputs from its original input up to $m$. Let us consider
$m=1\ldots1$ ($n$ 1s), and let us call $M$ with input $0\ldots0$ ($n$
0s). For each possible input of size $n$, $M'$ eventually reaches a
configuration where the tape only contains a sequence $y$, the cursor
is at the left boundary of $y$, and the state is $s'_A$ iff $M$
accepts $y$. Similarly, it eventually reaches a configuration where
the tape contains only sequence $y$, the cursor is at the left
boundary of $y$, and the state is $s'_R$ iff $M$ rejects $y$.

Let us note that $M'$ operates in polynomial space with respect to
$n$: for each sequence $y$ of length $n$, it executes $M$ for input
$y$, which runs in polynomial space with respect to $n$. The only
additional space used by $M'$ which was not used by $M$ is the space
used to keep the copy of the current input (whose size is $n$). Thus,
$M'$ runs in polynomial space.

Since property $A'$ applies to any Turing machine running in
polynomial space, it also applies to $M'$. Thus, if $A'$ and $P=NP$,
then we have an alternative method that lets us know, in polynomial
time, whether the original Turing machine $M$ accepts {\it any} input
$y$ of size $n$. We just have to find the well-chained
polynomial-size polynomial-execution-time function $f$ for machine
$M'$ and input $0\ldots 0$ as described in Section~\ref{sec:c}, and
next call $f$ for a configuration denoting that $y$ is on the tape,
the cursor is at its left boundary, and the state is $s'_A$. We have
that $f$ returns $1$ iff $M$ accepts $y$. The difference of this
method with respect to the method given in Section~\ref{sec:basic} is
that, after $f$ is found, we can use it for {\it all} inputs of the
selected size, that is, we do not need to find a {\it different}
function $f$ for each input of that size.

Let us note that, even if $P\neq NP$, $A'$ implies that such a
function $f$ exists for each size $n$. That is, for any $PSPACE$
problem, if $A'$ holds then there {\it exists} a fast and small
computation device that solves the problem {\it for all inputs of a
given size $n$}, even if $P\neq NP$. Provided that we already had
$f$, we could use it to quickly solve any instance of size~$n$.

\section{Conclusions}\label{sec:conclusions}

In this draft we have presented some properties that imply $P=NP
\Rightarrow P=PSPACE$. As far as we known, neither $A'$ implies $B$
nor the other way around, so the failure of one of them would not
imply the failure of the other.

If our current ideas to prove $A'$ and $B$ remain stuck and we lack
new ideas to do so, our next step will be performing the experiment
about $B$ which was mentioned in Section~\ref{sec:expB}.

\subsection*{Acknowledgements} I thank Javier Rodríguez-Laguna,
Fernando Rosa, and Fernando Rubio for their valuable comments on this
draft.

\bibliographystyle{plain}
\bibliography{concu}

\end{document}